\documentclass[%
 aps,
 prb,
 amsmath,amssymb,
 showpacs,
 reprint,%
]{revtex4-2}

\usepackage{graphicx}
\usepackage{dcolumn}
\usepackage{bm}
\usepackage{xcolor} 
\usepackage{verbatim}

\begin{document}

\title{Derivation of $d$-wave symmetry in real-space superconductors}

\author{Pavel Kornilovitch}
 \email{pavel.kornilovich@gmail.com}
 \affiliation{Department of Physics, Oregon State University, Corvallis, Oregon 97331, USA} 

\date{\today}  

\begin{abstract}

A simple derivation of $d$-wave order parameter within the real-space pairing and Bose-Einstein condensation mechanism of superconductivity is given. The two key ingredients are a short-range attraction between carriers and hole-like hopping between equal-energy pair configurations.       

\end{abstract}



\maketitle

\section{\label{DT:sec:one}
Introduction
}

In Ref.~\cite{Kornilovitch2025}, it was shown that $d$-symmetric hole pairs can form with zero attractive strength despite the presence of a Hubbard repulsion and a fully developed {\em three-dimensional} kinetic energy. For low pair density, it immediately implies a $d$-symmetric macroscopic order parameter since the latter is proportional to the pair wave function \cite{Bogoliubov1970}. Thus, $d$ wave can be explained within the real-space pairing mechanism, also known as bipolaron superconductivity \cite{Alexandrov1981,Alexandrov1994} or Bose-Einstein Condensation (BEC) superconductivity \cite{Ogg1946,Schafroth1954a,Schafroth1954b,Schafroth1957,Blatt1964,Eagles1969,Micnas1990}, under quite general assumptions. An effective real-space attractive potential may originate from long-range phonon interactions \cite{Froehlich1954,Verbist1991}, short-range lattice effects \cite{Zhang1991,Catlow1998,Edwards2023}, Jahn-Teller distortions \cite{Bednorz1988,Mihailovic2001,Kabanov2002}, spin fluctuations \cite{Scalapino2012}, or other sources \cite{Micnas1990}.           

Feedback on Ref.~\cite{Kornilovitch2025} suggested that these conclusions may further benefit from an elementary derivation of the key result, i.e., that the ground state of two holes has a $d$-symmetric wave function. Presenting this derivation is the purpose of this pedagogical note. While not intended for peer-reviewed publication, it complements Ref.~\cite{Kornilovitch2025} in making the main argument as transparent as possible.        

We note that from a symmetry viewpoint, solving a two-particle lattice problem is equivalent to solving a one-particle lattice problem in a static potential. Since the lattice is not Galilean-invariant, center-of-mass pair motion complicates the picture by mixing symmetries and changing dynamic pair properties \cite{Kornilovitch2023}. But in the ground state, i.e., at zero pair momentum, the symmetry of relative pair motion is the same as that of a single particle in a static potential. 

We limit consideration to two-dimensional square lattices because it is the simplest system where the difference between $s$ and $d$ symmetries can be observed. Physical reasoning presented herein applies equally well to more complex three-dimensional lattices such as the simple tetragonal lattice \cite{Adebanjo2024} or the body-centered tetragonal lattice \cite{Kornilovitch2025}.

\section{\label{DT:sec:two}
Models
}

\begin{figure*}[t]
\includegraphics[width=0.90\textwidth]{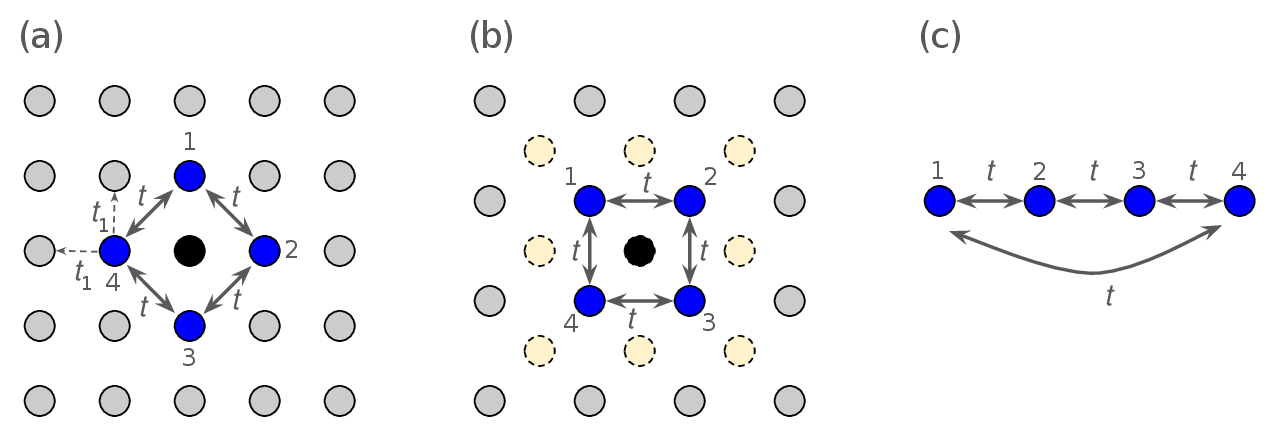}
\caption{(a) Square $UV$ model with second-nearest-neighbor hopping $t$. The black circle in the middle is the location of a spin-down fermion that serves as a source of attraction for the spin-up fermion on the four nearest-neighbor sites shown by blue circles. The four sites have the same energy and are connected by second-nearest neighbor hopping $t$. Notice that {\em first}-nearest neighbor hopping $t_1$ takes the spin-up fermion away from the lowest-energy configurations, as shown by dashed arrows. Hopping to double occupancy, i.e., when the spin-up fermion hops from a blue to the black circle, is prohibited by Hubbard repulsion $U$. Therefore, first-nearest-neighbor hopping can be disregarded in analysis of pair orbital symmetry in the strong-coupling limit, $V, U \gg t_{1} , t$. (b) Two layers of a body-centered tetragonal $UV$ model with {\em inter-layer} attraction. Both layers are two-dimensional square lattices, one is shown by circles with solid edges and another by circles with dashed edges. The layers are shifted relative to each other in the $z$ direction as well as by half the square diagonal in the $xy$ plane. The spin-down fermion resides on the central site of the ``dashed'' layer shown by the black circle. It serves as a source of attraction for the spin-up fermion that moves within the ``solid'' layer. The four configurations with lowest energies are shown by blue circles. In a full three-dimensional model \cite{Kornilovitch2025}, the fermions can also hop between layers but this hopping is irrelevant because it leads to high-energy configurations. (c) An equivalent four-site lattice chain with periodic boundary conditions.} 
\label{DT:fig:one}
\end{figure*}

We consider two concrete models depicted in Fig.~\ref{DT:fig:one}(a) and (b). The first model (a) is a square $UV$ model with nearest-neighbor attraction $V$ and {\em second}-nearest neighbor hopping $t$. A spin-down partner fermion resides on the central site shown by the black circle. The spin-down fermion is the source of static attraction for a spin-up fermion when the latter occupies the four sites surrounding the central site. Those four sites are shown by blue circles. Although first-nearest-neighbor hopping $t_1$ is more prominent on the square lattice, any first-nearest-neighbor hopping event takes the spin-up fermion away from a low-energy configuration, and as such can be disregarded in qualitative analysis of pair symmetry. In contrast, second-nearest-neighbor hopping $t$ connects sites of the same energy and must be included. In the context of $d$ symmetry, such square $UV$ model with nearest-neighbor attraction and second-nearest-neighbor hopping was considered by Blaer, Ren, and Tchernyshyov \cite{Blaer1997} and by Bak and Micnas \cite{Bak1999} some time ago, and more recently in Ref.~\cite{Adebanjo2024}.         
  
In the second model, Fig.~\ref{DT:fig:one}(b), {\em two} square lattices are offset in $z$ direction and shifted relative to each other in $xy$ plane by half of a square diagonal. Such an arrangement mimics out-of-plane attraction in the body-centered tetragonal $UV$ model introduced in Ref.~\cite{Kornilovitch2025}. The spin-down fermion resides on the upper layer shown by yellow circles with dashed edges and serves as a source of attraction for the spin-up fermion moving in the bottom layer shown by gray circles. In contrast with the first model, it is the {\em first} nearest-neighbor hopping $t$ that connects the four resonant pair states.   

We point out that for the purposes of this note it is sufficient to analyze the problem in the strong-coupling limit, $V, U \gg t$. It has been shown \cite{Kornilovitch2025} that the main conclusion holds for all couplings but the complexities of an exact solution would obscure the simple physical picture discussed here. It is therefore sufficient to keep only the four resonant pair configurations marked in Fig.~\ref{DT:fig:one} by numbers 1 through 4. In this approximation, both models become isomorphic to a four-site one-dimension chain shown in Fig.~\ref{DT:fig:one}(c). We have arrived at the following {\em one-particle} effective hopping model     
\begin{equation}
H = t \sum_{m , b = \pm 1 } c^{\dagger}_{ m + b } c_{ m } \: , 
\label{DT:eq:one}
\end{equation}
where $m = 1 , 2 , 3 , 4$ numbers the four configurations and $c$ are electron operators.

The sign of $t$ is now discussed. In model (b), intra-layer single-particle dispersion is defined solely by $t$, that is, $\varepsilon({\bf k}) = 2 t ( \cos{k_x} + \cos{k_y} )$. If $t < 0$, then the band minimum occurs at $k_x = k_y = 0$, the dispersion has a positive curvature, and carriers at low density are said to be electron-like. Four band maxima occur at $k_x = k_y = \pm \pi$, and the dispersion near the maxima has a negative curvature. If the band is almost filled, then effective low-density carriers are holes. An equivalent description of almost-filled band is the low-density limit of $\varepsilon({\bf k}) = 2 t ( \cos{k_x} + \cos{k_y} )$ but with $t > 0$. In this case, band minima occur at $k_x = k_y = \pm \pi$, and low-density carriers are hole-like. For these reasons, we will refer to $t < 0$ as ``electron-like'' hopping and to $t > 0$ as ``hole-like'' hopping when discussing the low-carrier-density limit.  

In model (a), one-particle dispersion is defined by both second-nearest-neighbor hopping $t$ and first-nearest-neighbor hopping $t_1$. Within a bound pair, however, it is the second-nearest-neighbor $t$ that matters. By analogy with model (b), we will be calling $t < 0$ electron-like hopping and $t > 0$ hole-like hopping also in model (a). Model (a) with hole-like second-nearest-neighbor hopping routinely occurs in studies of repulsive Hubbard model near half-filling, where a positive $t$ indicates the tendency of hole-like carriers \cite{Khait2023} to move through the lattice mostly diagonally.      

Finally, we note that electron-like and hole-like hoppings are connected by a particle-hole transformation. A hole is absence of an electron, hence    
\begin{equation}
p_{m} = c^{\dagger}_{ m } \: , 
\hspace{1.0cm}
p^{\dagger}_{m} = c_{ m } \: ,   
\label{DT:eq:two}
\end{equation}
where $p$ are hole operators. The Hamiltonian in Eq.~(\ref{DT:eq:one}) transforms as
\begin{equation}
H = t \sum_{m , b = \pm 1 } p_{ m + b } p^{\dagger}_{ m } = 
  - t \sum_{m , b = \pm 1 } p^{\dagger}_{ m + b } p_{ m }    \: , 
\label{DT:eq:three}
\end{equation}
where index shifts have been used. Thus, transition from electrons to holes or vice versa flips the sign of hopping. This is a mathematical justification of using negative $t$ for electrons and positive $t$ for holes.

\begin{figure*}[t]
\includegraphics[width=0.90\textwidth]{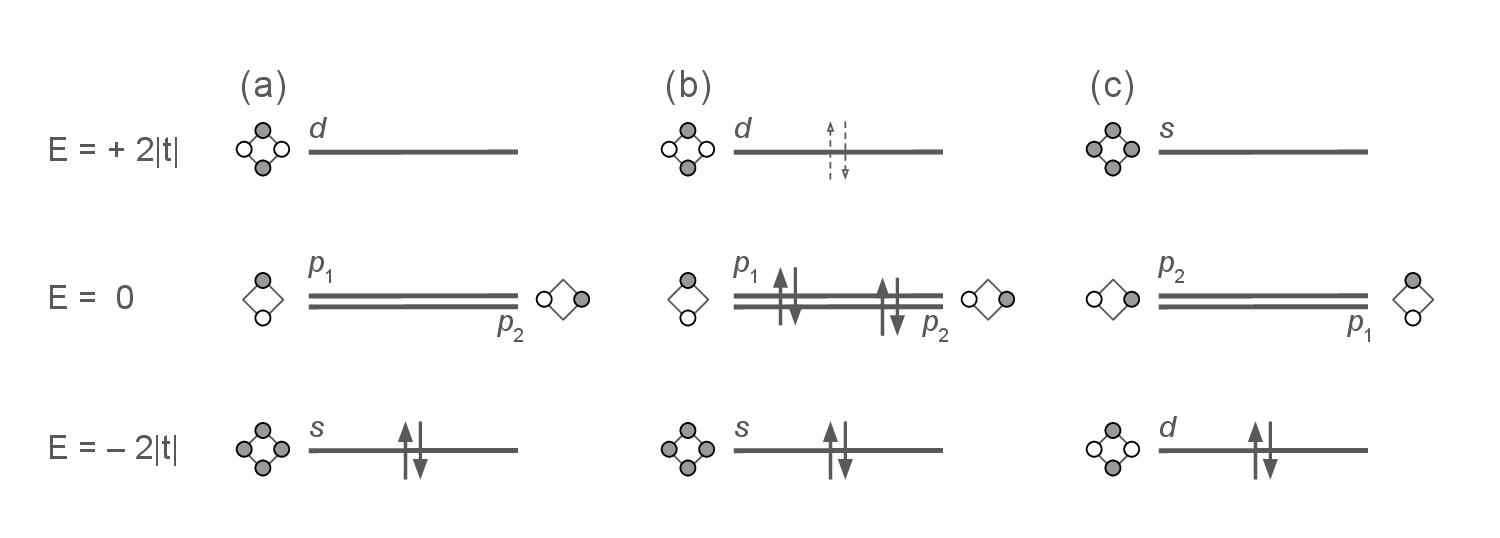}
\caption{Energy diagrams. There is always a $p$-symmetric doublet in the middle of the energy ladder. Wave function symmetries are shown next to the energy levels, dark-filled circles representing $\psi = + 1$ and light-filled circles representing $\psi = - 1$. (a) One electron pair in the electron picture, $t < 0$. One electron pair occupies the ground state and has $s$ symmetry. (b) One {\em hole} pair in the {\em electron} picture. The low energy states, one $s$ and two $p$'s, are occupied by electrons, leaving the top $d$ state unfilled. System's overall rotational symmetry is $d$. This can be interpreted as a sole hole pair (shown by dashed arrows) occupying the top $d$ level while all the lower levels are empty of holes. (c) Particle-hole transformation, Eq.~(\ref{DT:eq:two}), changes the sign of hopping, $- |t| \rightarrow + |t|$, flips the energy ladder, and replaces lack of a fermion with one's presence and vice versa. As a result, in the ground state, one hole pair occupies the lowest $d$-symmetrical state while the three upper states are empty. The overall system's symmetry is $d$. The physics behind diagrams (b) and (c) are identical.} 
\label{DT:fig:two}
\end{figure*}

\section{\label{DT:sec:three}
Energies and symmetries 
}

The Schr\"odinger equation for one-particle wave function $\psi_{m}$ and Hamiltonian (\ref{DT:eq:one}) reads
\begin{eqnarray}
t \psi_{2} + t \psi_{4} & = & E \psi_{1} 
\label{DT:eq:four} \\
t \psi_{1} + t \psi_{3} & = & E \psi_{2} 
\label{DT:eq:five} \\
t \psi_{2} + t \psi_{4} & = & E \psi_{3} 
\label{DT:eq:six}  \\
t \psi_{3} + t \psi_{1} & = & E \psi_{4} \: , 
\label{DT:eq:seven}  
\end{eqnarray}
where $E$ is counted from energy $V$ of a static configuration when the moving fermion occupies one of the attractive sites. There are two groups of states, depending on whether $E$ is zero or nonzero. 

\underline{$E = 0$.} From Eqs.~(\ref{DT:eq:four}) and (\ref{DT:eq:six}) it follows that $\psi_4 = - \psi_2$, and from Eqs.~(\ref{DT:eq:five}) and (\ref{DT:eq:seven}) it follows that $\psi_3 = - \psi_1$. Thus we have two orthogonal solutions that have $p$ symmetry with respect to plane rotations:
\begin{eqnarray}
p_1: \hspace{0.3cm} & & E = 0 ; \; \psi_{1} = 1 , \; \psi_{2} = 0 , \; \psi_{3} = -1 , \; \psi_{4} = 0  ; 
\label{DT:eq:eight} \\
p_2: \hspace{0.3cm} & & E = 0 ; \; \psi_{1} = 0 , \; \psi_{2} = 1 , \; \psi_{3} =  0 , \; \psi_{4} = -1 .  
\label{DT:eq:nine}  
\end{eqnarray}
Since these {\em coordinate} wave functions are anti-symmetric with respect to space inversion (that is with respect to fermion exchange when considering two particle), these two pair states are spin-triplets. 
  
\underline{$E \neq 0$.} In this case, from Eqs.~(\ref{DT:eq:four}) and (\ref{DT:eq:six}) it follows that $\psi_1 = \psi_3$, and from Eqs.~(\ref{DT:eq:five}) and (\ref{DT:eq:seven}) it follows that $\psi_2 = \psi_4$. As a result, four equations are reduced to two:
\begin{eqnarray}
2 t \psi_{2} & = & E \psi_{1} 
\label{DT:eq:ten} \\
2 t \psi_{1} & = & E \psi_{2} \: . 
\label{DT:eq:eleven} 
\end{eqnarray}
This system of equations has two eigenvalues, $E = \pm 2t$, but the symmetry of the ground state depends on the sign of hopping $t$. We begin with a negative $t$, i.e., with electron-like hopping.  

\underline{Electron-like hopping, $t < 0$.} Let us write for clarity $t = - |t|$. The system, Eqs.~(\ref{DT:eq:ten})-(\ref{DT:eq:eleven}), becomes:
\begin{eqnarray}
- 2 |t| \, \psi_{2} & = & E \psi_{1} 
\label{DT:eq:twelve} \\
- 2 |t| \, \psi_{1} & = & E \psi_{2} \: . 
\label{DT:eq:thirteen} 
\end{eqnarray}
Eigenvalues of this system are $E = \pm 2 |t|$. In the ground state, $E = - 2|t|$, and Eqs.~(\ref{DT:eq:twelve}) and (\ref{DT:eq:thirteen}) both yield $\psi_{1} = \psi_{2}$. Together with $\psi_{3} = \psi_{1}$ and $\psi_{4} = \psi_{2}$ established earlier, this implies $s$ rotational symmetry of the wave function. The second state has energy $E = + 2|t|$, and both Eqs.~(\ref{DT:eq:twelve}) and (\ref{DT:eq:thirteen}) yield $\psi_{2} = - \psi_{1}$, which corresponds to $d$ symmetry. Thus, we have
\begin{eqnarray}
s: \hspace{0.1cm} & & E = - 2 |t| ; \: \psi_{1} = 1 , \: \psi_{2} =  1 , \: \psi_{3} = 1 , \: \psi_{4} =  1 ; 
\label{DT:eq:fourteen} \\
d: \hspace{0.1cm} & & E = + 2 |t| ; \: \psi_{1} = 1 , \: \psi_{2} = -1 , \: \psi_{3} = 1 , \: \psi_{4} = -1 . 
\makebox[0.5cm]{} 
\label{DT:eq:fifteen}  
\end{eqnarray}
Since these coordinate wave functions are symmetric with respect to space inversion (that is with respect to fermion exchange when considering two particle), these two pair states are spin-singlets. A full energy diagram for the electron-like hopping is shown in Fig.~\ref{DT:fig:two}(a). 

From physical intuition, one expects that {\em holes} in the electron picture occupy highest energy levels. To create a single hole pair, one should fill all pair states with electrons from the bottom, but leave one state unfilled. This is shown in Fig.~\ref{DT:fig:two}(b). Since the highest pair state in the electron picture has $d$ symmetry, we expect the sole hole pair state to be $d$-symmetric.

\underline{Hole-like hopping, $t > 0$.} The above physical picture is confirmed mathematically, as illustrated in Fig.~\ref{DT:fig:two}(c). As discussed in Sec.~\ref{DT:sec:two}, particle-hole transformation, Eq.~(\ref{DT:eq:two}), flips the sign of hopping. Transitioning from the electron to hole picture, the hopping amplitude changes from negative to positive. For clarity, we continue to write the modulus sign, i.e., $t = + |t|$. Instead of Eqs.~(\ref{DT:eq:twelve}) and (\ref{DT:eq:thirteen}), we have 
\begin{eqnarray}
+ 2 |t| \, \psi_{2} & = & E \psi_{1} 
\label{DT:eq:sixteen} \\
+ 2 |t| \, \psi_{1} & = & E \psi_{2} \: . 
\label{DT:eq:seventeen} 
\end{eqnarray}
Eigenvalues of this system are still $E = \pm 2 |t|$ and the ground-state energy is still $E = - 2 |t|$. However, now Eqs.~(\ref{DT:eq:sixteen}) and (\ref{DT:eq:seventeen}) both yield $\psi_{2} = - \psi_{1}$ for the ground state, which corresponds to $d$ symmetry. The higher-energy state has $E = + 2 |t|$ and $\psi_{2} = + \psi_{1}$, which corresponds to $s$ symmetry. The two states are:
\begin{eqnarray}
d: \hspace{0.1cm} & & E = - 2 |t| ; \: \psi_{1} = 1 , \: \psi_{2} = -1 , \: \psi_{3} = 1 , \: \psi_{4} = -1 ; 
\makebox[0.5cm]{} 
\label{DT:eq:eighteen} \\
s: \hspace{0.1cm} & & E = + 2 |t| ; \: \psi_{1} = 1 , \: \psi_{2} =  1 , \: \psi_{3} = 1 , \: \psi_{4} =  1 . 
\label{DT:eq:nineteen}  
\end{eqnarray}
Thus, the particle-hole transformation has flipped the energy diagram: the highest state in the electron picture became the lowest state in the hole picture. The physics is unchanged: a single {\em hole} pair is $d$-symmetrical, regardless of whether electrons or holes are used to describe it.

\section{\label{DT:sec:four}
Connection with a macroscopic order parameter 
}

At a low carrier density, electron or hole, macroscopic superconductor order parameter $\Delta$ is proportional to ground-state pair wave function $\psi_0$:
\begin{equation}
\Delta( {\bf r}_1 , {\bf r}_2 ) = \sqrt{ \frac{N_0}{\Omega} } \, \psi_{0}( {\bf r}_1 - {\bf r}_2 ) \: , 
\label{DT:eq:twenty}
\end{equation}
where $N_0$ is the number of pairs in the condensate and $\Omega$ is the system's volume. This relation was derived by Bogoliubov in Ref.~\cite{Bogoliubov1970} and reproduced in Ref.~\cite{Kornilovitch2025}. A practical application of Eq.~(\ref{DT:eq:twenty}) is that the orbital symmetries of $\Delta$ and $\psi_0$ are the same. Thus, a $d$-symmetrical bound state of two holes implies a $d$ macroscopic order parameter in the thermodynamic limit.   

In closing, we mention that all the qualitative results related to pair symmetry derived in this note in the strong-coupling limit remain valid at all couplings, as confirmed by rigorous solution of the two-particle Schr\"odinger equation \cite{Kornilovitch2025}.





\end{document}